\documentclass{PoS}

\usepackage{wrapfig}              
\usepackage{adjustbox}          

\title{Microscopic Theory of Nuclear Fission}

\ShortTitle{Microscopic Theory of Nuclear Fission}

\author{\speaker{Aurel Bulgac}\thanks{The slides of the presentation can be downloaded from 
{http://faculty.washington.edu/bulgac/Pu240/ }.}\\
        Department of Physics, University of Washington., Seattle, WA 98195-1560, USA\\
        E-mail: \email{bulgac@uw.edu}}

\author{Shi Jin\\
        Department of Physics, University of Washington., Seattle, WA 98195-1560, USA\\
        E-mail: \email{js1412@uw.edu}}
        
\author{Piotr Magierski\\
        Faculty of Physics, Warsaw University of Technology, ulica Koszykowa 75, 00-662 Warsaw, POLAND\\
        E-mail: \email{Piotr.Magierski@if.pw.edu.pl}}      
        
\author{Kenneth J. Roche\\
        Pacific Northwest National Laboratory, Richland, WA 99352, USA\\
        E-mail:\email{Kenneth.Roche@pnnl.gov}}
         
\author{Ionel Stetcu\\
            Theoretical Division, Los Alamos National Laboratory, Los Alamos, NM 87545, USA \\    
            E-mail:\email{stetcu@lanl.gov}}

          \abstract{We describe the fission dynamics of $^{240}$Pu
            within an implementation of the Density Functional Theory
            (DFT) extended to superfluid systems and real-time
            dynamics.  We demonstrate the critical role played by the
            pairing correlations, which even though are not the
            driving force in this complex dynamics, are providing the
            essential lubricant, without which the nuclear shape
            evolution would come to a screeching halt.  The evolution
            is found to be much slower than previously expected in
            this fully non-adiabatic treatment of nuclear dynamics,
            where there are no symmetry restrictions and all
            collective degrees of freedom (CDOF) are allowed to
            participate in the dynamics.  }

\FullConference{The 26th International Nuclear Physics Conference\\
		11-16 September, 2016\\
		Adelaide, Australia}

\begin{document}

\begin{wrapfigure}{r}{9cm}
\includegraphics[clip,width=9cm]{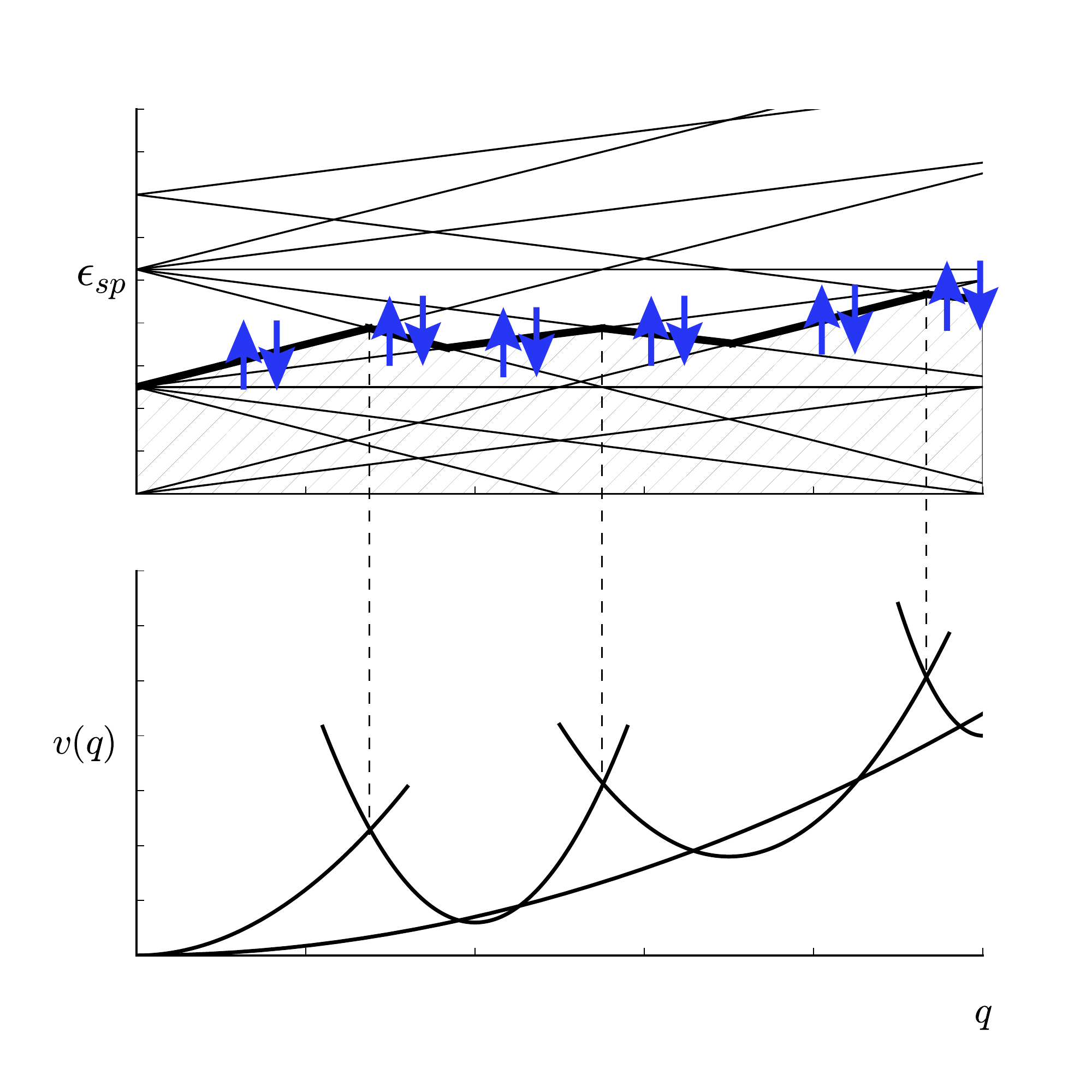}
\caption{ \label{fig:ab2}    The qualitative evolution of the single-particle levels and of the total nuclear energy (lower panel) as a function of  
nuclear deformation~\cite{PhysRev_1953r, Bertsch:1980}. The Fermi level is shown with a thick line.}
\end{wrapfigure}

Immediately after the epochal discovery of induced nuclear fission by
Hahn and Strassmann~\cite{Natur_1939r} Meitner and
Frisch~\cite{Nature_1939r}, Bohr and
Wheeler~\cite{Nature_1939r1,PhysRev_1939r} recognized that the main
driving force leading to fission is the profile of the deformation
nuclear energy arising form the competition between the nuclear
surface and the Coulomb energies. The reasoning was based on a
classical liquid charged drop model of a nucleus and the role of
quantum mechanics started to became clear only with time. It was
however soon realized that in nuclei nucleons form shells and behave
in many instances as independent particles, like electrons in atoms,
or in other words that the nucleons live on quantized
orbits~\cite{Mayer,Jensen}, and that the spin-orbit interaction plays
a critical role in the formation of the nuclear shells.  Hill and
Wheeler~\cite{PhysRev_1953r} were apparently the first to appreciate
how the liquid drop deformation energy emerges from a quantum
mechanical approach based on considering the quantized single-particle
motion of nucleons in a slowly deforming potential well. The liquid
drop potential deformation energy in their approach in the first
approximation was an envelope of many intersecting parabolas, due to
single-particle level crossings, see Figure \ref{fig:ab2}. At
single-particle level crossings of the last occupied level nucleons jump from one level to another, in order to maintain the
sphericity of the Femi sphere. If a nucleus elongates on the way to
scission into two fragments, without such a redistribution of nucleons
at the Fermi level, the Fermi sphere would become oblate, while the
spatial shape of the nucleus becomes prolate, and that would lead to a
volume excitation energy of the nucleus. In the case of nuclei,  
which are saturating systems with a surface tension, while
deforming by changing the shape of their surface only and while maintaining
constant their volume, only the Coulomb and the surface
contributions to the total energy changes.
Each single-particle level is typically double degenerate, due to
Kramers degeneracy, and nucleons would have to jump in pairs,
otherwise the nuclear shape evolution towards scission would be
hindered~\cite{Bertsch:1980,PhysRevLett_1997r}. Pairing interaction,
which in spite of being relatively weak in nuclei, is very effective
of promoting simultaneously two nucleons from time-reverse orbits into
other time-reverse orbits and thus it greatly facilitates the
evolution of the nuclear shape towards scission.

It was established later that single-particle level bunching exist in
nuclear systems not only in the case of spherical nuclei (as in the
case of atoms), but also in deformed and highly deformed nuclei. At
first this phenomenon was experimentally observed in the case of
fission isomers at very large
elongations~\cite{Polikanov,RMP_1972r,RMP_1980r} and subsequently in
the case of superdeformed nuclei~\cite{Twin}.  The existence of
nucleonic shells at large deformations results in a potential energy
deformation surface with significant maxima and minima, which are
otherwise absent in the case of a classical charged liquid drop. The
level crossings lead to a potential energy surface which appears quite
rough, even though it can be smoothed out in the presence of pairing
correlations, which results in avoided level crossings. The nuclear
deformation potential energy surface appears in the end to have a rather
complicated structure.   The gross behavior is determined by the surface
and Coulomb energy and resembles the deformation energy of a charged
liquid drop and that is the main driving force leading to
fission. Because nuclei are relatively small quantum systems made of a
bit more than a couple hundred fermions, which to a large extend
behave as being independent, a rather rich shell structure exists,
even for large deformations. This shell structure imprints on the
overall charged liquid drop energy hills and
valleys~\cite{RMP_1972r,RMP_1980r}. On the way to the scission
configuration nucleons have to perform a large number of
redistributions between the single-particle levels crossing at the
Fermi level, in order to maintain the spherical symmetry of their
local momentum distribution or of the Fermi sphere. Overall, the
deformation potential energy surface acquires a profile somewhat
similar to that of an uneven mountain, with little hills and valleys
and covered by trees, and the evolution of the nuclear shape is in the
end similar to the erratic motion of a pinball, not straight down the
hill, but rather left and right, bouncing (mostly elastically) from
the many obstacles on the way to the bottom of the valley, where the
pinball breaks up. At the last stages of this complex nuclear shape
evolution the independent character of the nucleons inside nuclei
plays again a critical role, magic closed shells control the nuclear
shape evolution.  As our simulations demonstrate\cite{Bulgac:2016},
the nucleus separates typically into two fission fragments, one bigger
and the other somewhat smaller. The larger fragment fragment has
properties very similar to the energetically very stable double-magic
$^{132}$Sn, emerges with an almost spherical shape, while the lighter
fragment at the scission emerges into an elongated shape, with a ratio
of the major to minor axes close to 3/2.

Overall the fission dynamics is a very complex process, which still
did not reach a full microscopic description~\cite{arxiv_1511r}, in
spite of almost eight decades of effort. In contrast the
superconductivity, another remarkable quantum many-body phenomenon,
required less than five decades to reach a microscopic
understanding. Several reasons prohibited so far the formulation of a
microscopic theory of fission (as opposed to phenomenological models),
capable to produce results comparable to observations without
introducing uncontrollable approximations, parameter fitting, and
based on microscopic input.  Two major relatively recent developments
proved to be critical and created the conditions for the formulation
of a microscopic theory of fission. The first element was the
extension of the Density Functional Theory (DFT) to superfluid fermion
systems, and extension in the spirit of the Kohn and Sham~\cite{ks}
Local Density Approximation (LDA) from normal systems to superfluid
systems~\cite{PRL__2002,PRL__2003a,ARNPS__2013}, the Superfluid Local
Density Approximation (SLDA). A second major development was the
mergence of powerful supercomputers capable of handling a
time-dependent DFT (TDDFT) approach to nuclear fission.

Since fission dynamics is a truly non-stationary phenomenon a further
extension was needed to its Time-Dependent version~\cite{ARNPS__2013}
and this approach was dubbed the Time-Dependent Superfluid Local
Density Approximation (TDSLDA). According to the theorem of Hohenberg
and Kohn~\cite{hk} there is a one-to-one correspondence between the
full ground state many-body wave function of a fermion system and the
one-body density matrix: $\Psi(x_1,\ldots,x_N) \leftrightarrow n({\bf
r}),$ which directly leads to the fact that an energy density
functional (EDF) exists: $E_{gs} = \langle \Psi [n] | H |\Psi [n]
\rangle \equiv \int d^3r \varepsilon ({\bf r}).  $
\begin{wrapfigure}{r}{0.55\textwidth}
\includegraphics[clip,width=0.55\textwidth]{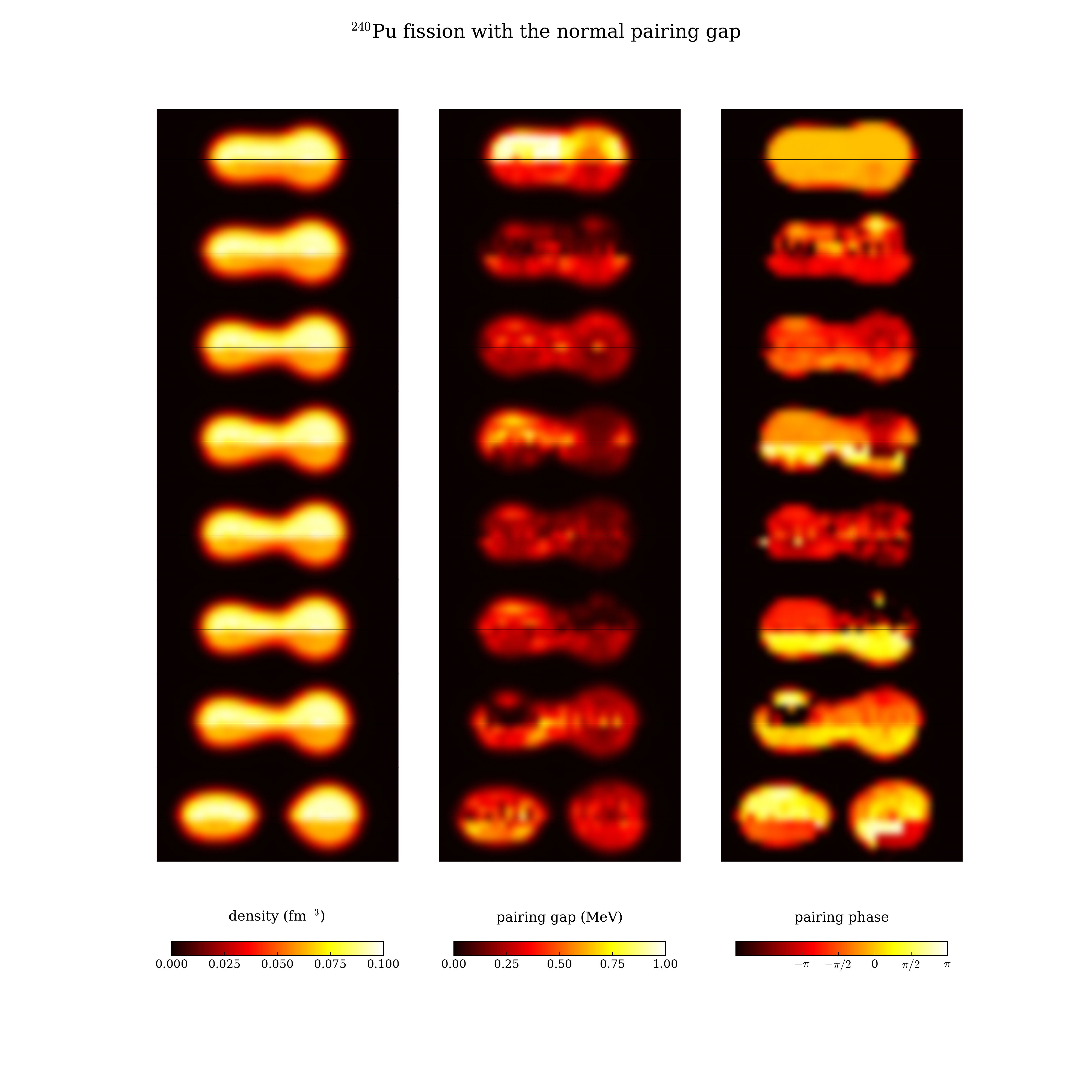}
\caption{ \label{fig:ab3}   Induced fission of $^{240}$Pu with normal
pairing strength last about 14,000 fm/c from saddle-to-scission. The
columns show sequential frames of the density (first column),
magnitude of the pairing field (second column), and the phase of the
pairing field (third column).  In each frame the upper/lower part of
each frame shows the neutron/proton density, the magnitude of
neutron/proton pairing fields, and of the phase of the pairing field
respectively~\cite{Bulgac:2016}. At scission the heavy fragment is on
the right and emerges almost spherical, while the light fragment is
highly deformed with the ration of the axes $\approx 3/2$.}
\end{wrapfigure}
The extension of these statements to both excited states and
time-dependent phenomena has been performed for quite some time now
and many aspects are well documented in
monographs~\cite{monograph1,monograph2}. Unfortunately so far no
recipes have been produced on how to generate the energy density
functional and only semi-phenomenological solutions, quite accurate
though, have been suggested.  For this reason many still in the
nuclear physics community are still leveling an unwarranted criticism
at the DFT. DFT, similarly to the Schr\"{o}dinger equation in quantum
mechanics, provides the theoretical framework within which one has to
attack a variety of quantum many-body problems. In the Schr\"{o}dinger
equation one has to provide the potential, which in most cases we know
only approximately, with various degrees of accuracy. The same is true
in the case of DFT, the energy density functional is known only with
some degree of accuracy. One would not stop using the Schr\"{o}dinger
equation if one would know the potential only approximately and
instead would revert to some alternative methods.  What are the
minimal requirements a nuclear EDF (NEDF) has to meet in order to be
used the fission dynamics within a TDDFT approach? Apart from the
usual constraints (translational and rotational invariance, Galilean
invariance, isospin symmetry, parity,etc.) it has to describe accurately
the saturation properties on nuclear matter, have a correct surface
tension and accurate spin-orbit interaction, and accurate pairing
properties.  Saturation properties, surface energy and Coulomb energy
are needed to describe nuclear fission at the liquid drop model. The
spin-orbit interaction and accurate pairing energies are needed in
order to describe correctly the shell-corrections and the fact that
the emerging heavy fission fragment in the fission of actinide has
properties very close to the double-magic $^{132}$Sn. Pairing
correlations are also critical in order to have an efficient mechanism
to maintain the sphericity of the Fermi surface while the shape of the
mother nucleus evolves from a compact form up to scission. In the
absence of an efficient mechanism, which would redistribute the
nucleon pairs at the level-crossings occurring at the Fermi level, see
Figure~\ref{fig:ab2}, a nucleus would fail to fission, unless excited
to very large energies or elongated well beyond the our fission
barrier~\cite{PhysRevLett_1997r,arxiv_2015r,arxiv_2015r1}.

An extension of the DFT to a TDDFT of superfluid systems requires the
introduction of two new order parameters: the anomalous densities and
currents. Pairing interaction in nuclei is short-ranged and that
results in anomalous densities which converge very slowly with the
upper cutoff energy, which is of the order of 100 MeV. It can be shown
that if the pairing potential is local the anomalous density is
actually divergent~\cite{PRL__2002,PRL__2003a}.  Trying to eschew the
presence of a divergent anomalous density by using finite short-range
interactions makes the TDDFT approach practically impossible to
implement in practice. An interaction such as the very successful
Gogny interaction, has an phenomenological short-range, not resulting
from any microscopic input.  Moreover, a finite range interaction will
render the TDDFT equations into partial differential-integral
equations, which would require computations resources well beyond
exascale computers. And last, but not least, there is no fundamental
reason why the DFT equations have to be non-local in space in the case
of nuclear interactions. It suffices to say that that was no need for non-local equations  in
the case of the electronic systems, for which the Coulomb interaction
has an infinite range. We have used an NEDF based on the popular
phenomenological SLy4 interaction~\cite{NuclPhys_1998}, supplemented
with a very accurate pairing anomalous contribution~\cite{PRL__2003a}.

\begin{wrapfigure}{r}{0.55\textwidth}
\includegraphics[clip,width=0.55\textwidth]{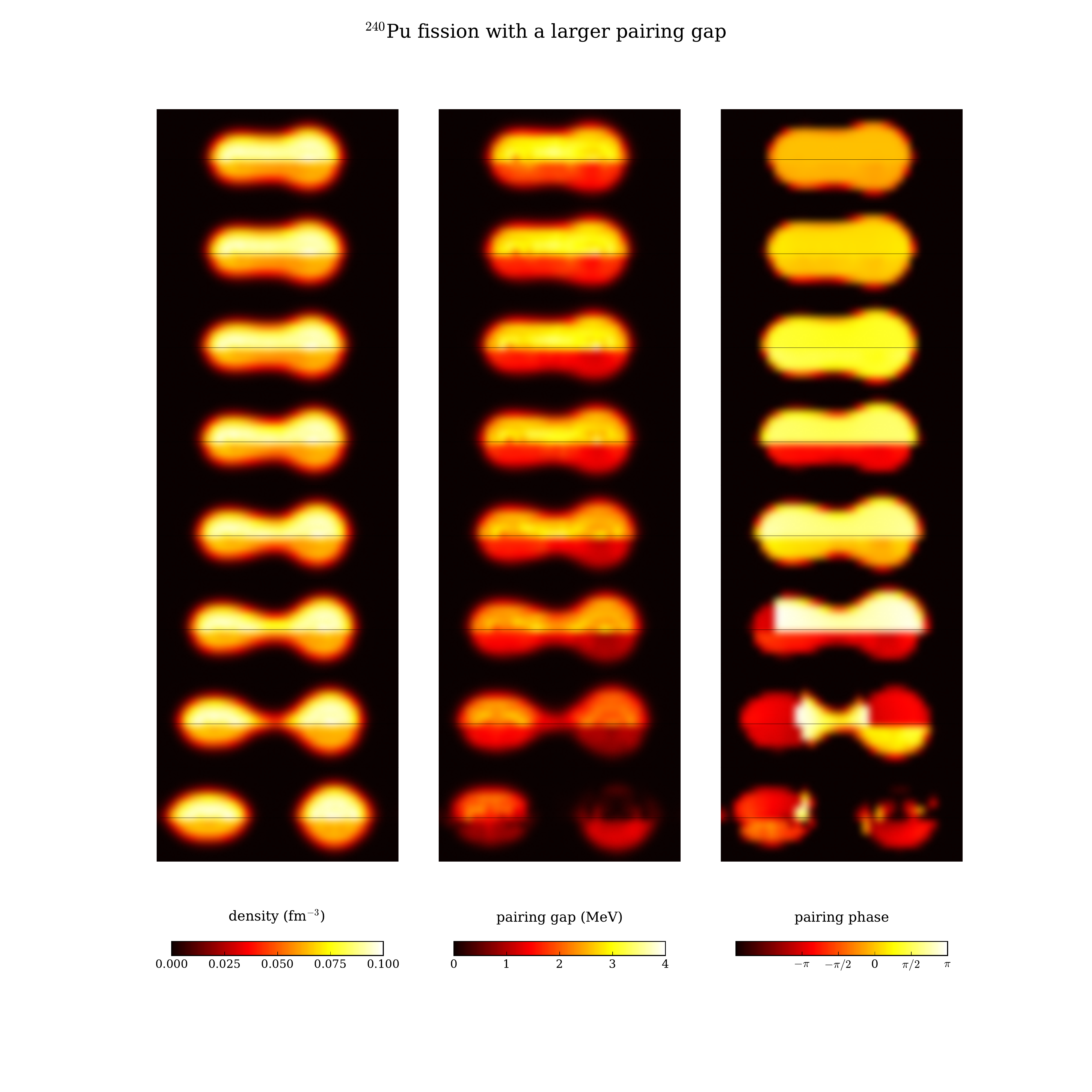}
\caption{ \label{fig:ab4}    Induced fission of $^{240}$Pu with
enhanced pairing strength last about 1,400 fm/c from
saddle-to-scission, thus about ten times faster than in the case of
normal pairing strength.}
\end{wrapfigure}

The emerging TDSLDA equations appear by design formally as TDHFB
equations with local meanfield and pairing potentials. One has to
remember that unlike TDHF or TDHFB equations, which are derived
following specific approximations, the TDSLDA equations have an exact
theoretical structure, and the only approximation is in the actual NEDF used, in our case
a phenomenological one. The SLy4 NEDF provides quite an accurate
description of the saturation properties of nuclear matter, of the
surface properties of nuclei, and leads to a correct reproduction of
the magic numbers, thus has a reasonably accurate spin-orbit
contribution to the NEDF.  The pairing part of the NEDF used by us is
also quite accurate~\cite{PRL__2003a}. The time-dependent equations of
the TDSLDA are discretized on a spatial lattice in a box large enough
to contain both the initial mother nucleus as well as the separated
fission fragments immediately after scission. The lattice constant
corresponds to a momentum cutoff of $\approx 500$ MeV/c, which is
definitely large enough to describe accurately a large class of
low-energy nuclear phenomena. Each single particle wave functions has
four components: $u_{n\uparrow}({\bf r}), u_{n\downarrow},
v_{n\uparrow}({\bf r}), v_{n\downarrow}({\bf r})$, where $n$ runs over
proton and neutron quasiparticle states. The total number of partial 
differential equations varies, depending on the size of the
simulation box and the lattice constant, and runs from tens of thousands
to hundreds of thousands.  This very large number of coupled,
non-linear time-dependent 3D partial differential equations and the
very large number of time-steps required to complete the full
evolution of the system explains why such a problem could not have
been attacked numerically less than a decade ago.  We initialize the
fission nucleus to a state very close to the outer fission barrier and
let the system evolve until scission. No restriction of any type are
imposed on the dynamics and at all times the meanfield and the pairing
potentials are determined by the instantaneous nucleon densities, and
in this sense the dynamics is selfconsistent.

The most remarkable aspect of our work was that the chosen nucleus $^{240}$Pu,
with an excitation energy of about 8 MeV corresponding to the induced
fission $^{239}$Pu(n,f) with a neutron with an impinging kinetic
energy of about 1.5 MeV reached the scission configuration and
separated into two unequal fragments, see Figure \ref{fig:ab3}. The
average atomic mass $A_L\approx 105.3$, neutron number $N_L\approx
63.5$, and charge $Z_L\approx 39.7$ of the light fragment obtained in
simulations compare surprisingly well with the systematic data
$A^S_L\approx 100.6$, $N^S_L\approx 61$, and $Z_L^S\approx39.7$,
particularly considering that no effort or fitting was made. The light
fragment emerges very deformed at scission, with the shape of an
axially symmetric ellipsoid with the ratio of the major to the minor
axes close to 3/2. The deformation energy of the light fission
fragment is eventually converted into internal excitation energy and
as a result most of the excitation energy resides in the light fission
fragment. the total number of post-scission neutron emitted is
estimated between 2-3 in reasonable agreement with experiment. The
total kinetic energy of the fission fragments we obtain is 181.6 MeV
to be compared with the value obtained from systematics 177.3 MeV. The
time form the outer saddle-to-scission is surprisingly very large, of
the order of 10,000 fm/c, which is about an order of magnitude larger
that any previous estimate within various phenomenological models. The
dynamics appears superficially as over damped, but in reality the
down-the-hill roll of the nucleus can be compared with the motion of
an electron in the Drude model of electric conduction. An electron
collides with various ions in the lattice and it is forced to move in
transversal directions to the electric field, and even though there is
no dissipation, the total kinetic+potential energy of the "electron"
is conserved and at any "height" the magnitude of the velocity in the
presence and in the absence of the "ions" is the same, the actual
length of the trajectory is significantly longer, which results in a
much longer time to reach the bottom. This is exactly what happens in
the case of the evolution of a fissioning nucleus while it rolls down
from the outer saddle-to-scission. As there are no constraints on the
dynamics (as usually practitioners enforce in semi-microscopic and
phenomenological models), all collective degrees of freedom are
allowed to participate. The fission fragments at the scission
configuration are rather cold, any excitation is present mostly in the
collective (shape and pairing) degrees of freedom, while the intrinsic
nucleonic degrees of freedom follow essentially an adiabatic
evolution.

The deformation potential energy surface however is still full of
local little hills and valleys, arising from the partially avoided
level crossing as a result of the pairing correlations. To demonstrate
how essential the role of the pairing correlations is in the fission
dynamics, in spite of the fact that in magnitude the contribute very
little, we performed a fission study where we artificially increased
the strength of the pairing gaps, see Figure \ref{fig:ab4}. The main
result of this is that the magnitude of the roughness of the
deformation potential energy surface is greatly reduced and the
deformation potential energy surface is thus almost smooth. A nucleus
now at the top of the barrier will start rolling down the hill towards
scission configuration straight down, without significant excursions
sideways. The time from saddle-to-scission in this case is a factor
ten smaller and almost identical to the time one would obtain in a
fully hydrodynamic approach of an ideal nuclear fluid~\cite{NEDF}. This
results agrees with the known behavior of well developed superfluid
systems, which at zero temperature behave as ideal or perfect
fluids. There is another remarkable difference between the dynamics
illustrated in Figures \ref{fig:ab3} and \ref{fig:ab4}. While the
evolution of the neutron and proton densities appear superficially
similar, the evolution of the pairing field is qualitatively
different. In Figure \ref{fig:ab3} the paring field is seen to
fluctuate quite a lot both in magnitude and in phase, a signature of a
not very well developed condensate. In Figure \ref{fig:ab4} however,
prior to scission the pairing fields of both neutrons and protons
hardly fluctuate either in magnitude or phase, and as in stationary
ground states, the phase is essentially uniform throughout the entire
nuclear system, which is a signature of a well defined adiabatic
evolution of the entire system. This is an example of the recently
described mechanism of phase-locking in the evolution of superfluid
systems~\cite{Bulgac:2017}, when the strength of the interaction
leading to superfluidity (both in Fermi and Bose systems) exceeds a
critical value. We can thus conclude that the pairing interaction, in
spite of being rather weak, plays a crucial role in the nuclear
large amplitude dynamics. In its absence a nuclear system would come to
screeching halt~\cite{arxiv_2015r,arxiv_2015r1}, as in the presence of
large static friction in classical systems.  

This work was supported in part by U.S. Department of Energy (DOE)
Grant No. DE-FG02-97ER41014, the Polish National Science Center (NCN)
under Contracts No. UMO-2013/08/A/ST3 /00708 and
No. UMO-2012/07/B/ST2 /03907.  IS gratefully acknowledges partial
support of the U.S. Department of Energy through an Early Career Award
of the LANL/LDRD Program.  Calculations have been performed at the
OLCF Titan and at NERSC Edison.  This research used
resources of the Oak Ridge Leadership Computing Facility, which is a
DOE Office of Science User Facility supported under Contract
DE-AC05-00OR22725 and 
used resources of the National Energy Research
Scientific computing Center, which is supported by the Office of
Science of the U.S. Department of Energy under Contract
No. DE-AC02-05CH11231.


\end{document}